# On a Catalogue of Metrics for Evaluating Commercial Cloud Services

Zheng Li
School of CS
NICTA and ANU
Canberra, Australia
Zheng.Li@nicta.com.au

Liam O'Brien
CSIRO eResearch
CSIRO and ANU
Canberra, Australia
Liam.OBrien@csiro.au

He Zhang
School of CSE
NICTA and UNSW
Sydney, Australia
He.Zhang@nicta.com.au

Rainbow Cai
School of CS
NICTA and ANU
Canberra, Australia
Rainbow.Cai@nicta.com.au

*Abstract—* **Given the continually increasing amount of commercial Cloud services in the market, evaluation of different services plays a significant role in cost-benefit analysis or decision making for choosing Cloud Computing. In particular, employing suitable metrics is essential in evaluation implementations. However, to the best of our knowledge, there is not any systematic discussion about metrics for evaluating Cloud services. By using the method of Systematic Literature Review (SLR), we have collected the de facto metrics adopted in the existing Cloud services evaluation work. The collected metrics were arranged following different Cloud service features to be evaluated, which essentially constructed an evaluation metrics catalogue, as shown in this paper. This metrics catalogue can be used to facilitate the future practice and research in the area of Cloud services evaluation. Moreover, considering metrics selection is a prerequisite of benchmark selection in evaluation implementations, this work also supplements the existing research in benchmarking the commercial Cloud services.**

*Keywords- Cloud Computing; Commercial Cloud Service; Cloud Services Evaluation; Evaluation Metrics; Catalogue*

## I. INTRODUCTION

Cloud Computing, as one of the most promising computing paradigms [1], has become increasingly accepted in industry. Correspondingly, more and more commercial Cloud services offered by an increasing number of providers are available in the market [2, 5]. Considering that customers have little knowledge and control over the precise nature of commercial Cloud services even in the "locked down" environment [3], evaluation of those services would be crucial for many purposes ranging from cost-benefit analysis for Cloud Computing adoption to decision making for Cloud provider selection.

When evaluating Cloud services, a set of suitable measurement criteria or metrics must be chosen. In fact, according to the rich research in the evaluation of traditional computer systems, the selection of metrics plays an essential role in evaluation implementations [32]. However, compared to the large amount of research effort into benchmarks for the Cloud [3, 4, 16, 21, 34, 45], to the best of our knowledge, there is not any systematic discussion about metrics for evaluating Cloud services yet. Considering that the metrics selection is one of the prerequisites of benchmark selection [31], we proposed to perform a comprehensive investigation into evaluation metrics in the Cloud Computing domain.

Unfortunately, in contrast with traditional computing systems, the Cloud nowadays is still chaos [56]. The most outstanding issue is that there is a lack of consensus of standard definition of Cloud Computing, which inevitably leads to market hype and also skepticism and confusion [28]. As a result, it is hard to point out the range of Cloud Computing and a full scope of metrics for evaluating different commercial Cloud services. Therefore, we decided to unfold the investigation along a regression manner. In other words, we tried to isolate the de facto evaluation metrics from the existing evaluation work to help understand the state-of-the-practice of the metrics used in Cloud services evaluation. When it comes to exploring the existing evaluation practices of Cloud services, we employed three constraints:

- This study focused on the evaluation of only commercial Cloud services, rather than that of private or academic Cloud services, to make our effort closer to industry's needs.
- This study concerned Infrastructure as a Service (IaaS) and Platform as a Service (PaaS) without considering Software as a Service (SaaS). Since SaaS with special functionalities is not used to further build individual business applications [21], the evaluation of various SaaS instances could require infinite and exclusive metrics that would be out of the scope of this investigation.
- This study only explored empirical evaluation practices in academic publications. There is no doubt that informal descriptions of Cloud services evaluation in blogs and technical websites can also provide highly relevant information. However, on the one hand, it is impossible to explore and collect useful data from different study sources all at once. On the other hand, the published evaluation reports can be viewed as typical and peer-reviewed representatives of the existing ad hoc evaluation practices.

Considering that the Systematic Literature Review (SLR) has been widely accepted as a standard and rigorous approach to evidence collection for investigating specific research questions [26, 27], we adopted the SLR method to identify, assess and synthesize the published primary studies of Cloud services evaluation. Due to the limit of space, the detailed SLR process is not elaborated in this paper [1]. Overall, we have identified 46 relevant primary studies covering six commercial Cloud providers, such as Amazon, GoGrid, Google, IBM, Microsoft, and Rackspace, from a set of popular digital publication databases (all the identified primary studies have been listed online for reference: http://www.mendeley.com/groups/1104801/slr4cloud/papers /). More than 500 evaluation metrics including duplications were finally extracted from the identified Cloud services evaluation studies.

This paper reports our investigation result. After removing duplications and differentiating metric types, the evaluation metrics were arranged according to different Cloud service features covering the following aspects: Performance, Economics, and Security. The arranged result essentially constructed a catalogue of metrics for evaluating commercial Cloud services. In turn, we can use this metrics catalogue to facilitate the Cloud services evaluation work, such as quickly looking up suitable evaluation metrics, identifying current research gap and future research opportunities, and developing sophisticated metrics based on the existing metrics.

The remainder of the paper is organized as follows. Section II arranges all the identified evaluation metrics under different Cloud service features. Section III introduces three scenarios of applying this metrics catalogue. Conclusions and some future work are discussed in Section IV.

II. THE METRICS FOR CLOUD SERVICES EVALUATION

It is clear that the choice of appropriate metrics depends on the service features to be evaluated [31]. Therefore, we naturally organized the identified evaluation metrics according to their corresponding Cloud service features. In detail, the evaluated features in the reviewed primary studies can be found scattered over three aspects of Cloud services (namely Performance, Economics [35], and Security) and their properties. Thus, we use the following three subsections to respectively introduce those identified metrics.

*A. Performance Evaluation Metrics*

In practice, an evaluated performance feature is usually represented by a combination of a physical property of Cloud services and its capacity, for example Communication Latency, or Storage Reliability. Therefore, we divide a performance feature into two parts: *Physical Property* part and *Capacity* part. Thus, all the elements of performance features identified from the aforementioned primary studies can be summarized as shown in Figure 1. The detailed explanations and descriptions of different performance feature elements have been clarified in our previous taxonomy work [57]. In particular, Scalability and Variability are also regarded as two elements in the *Capacity* part, while further distinguished from the other capacities, because they are inevitably reflected by the changes in the index of normal performance features.

Naturally, here we display the performance evaluation metrics mainly following the sequence of these performance elements. In addition, the evaluation metrics for overall performance of Cloud services are particularly listed. The metrics for evaluating Scalability and Variability are also separated respectively.

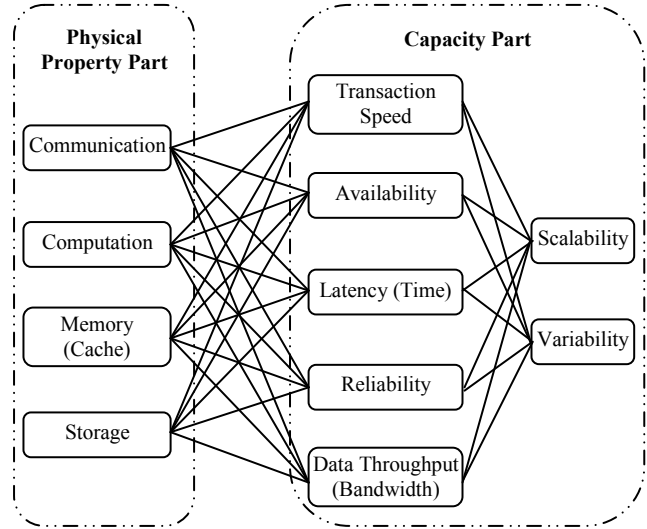

Figure 1. Performance features of Cloud services for evaluation.

*1) Communication Evaluation Metrics (cf. Table I):* Communication refers to the data/message transfer between internal service instances (or different Cloud services), or between external client and the Cloud. In particular, given the separate discussions about IP-level and MPI-message-level networking among public Clouds [e.g. 8], we also distinguished evaluation metrics between TCP/UDP/IP and MPI communications.

Brief descriptions of particular metrics in Table I:

- *Packet Loss Frequency vs. Probe Loss Rate:* Here we directly copied the names of these two metrics from [43]. Packet Loss Frequency is defined as the rate between loss_time_slot and total_time_slot, and Probe Lost Rate is defined as the rate between lost_probes and total_probes. Considering that the concept Availability is driven by the time lost while Reliability is driven by the number of failures [10], we can find that the former metric is for Communication Availability evaluation while the latter is for Communication Reliability.
- *Correlation between Total Runtime and Communication Time:* This metric is to observe a set of applications about their runtime and the amount of time they spend communicating in the Cloud. The trend of the correlation can be used to qualitatively discuss the influence of Communication on the applications running in the Cloud.

---
[1] The SLR report can be found online:
https://docs.google.com/open?id=0B9KzcoAAmi43LV9IaEgtNnVUenVXSy1FWTJKSzRsdw

TABLE I. COMMUNICATION EVALUATION METRICS

| Capacity | Metrics | Benchmark |
|---|---|---|
| Transaction Speed | Max Number of Transfer Sessions | SPECweb 2005 [22] |
| Availability | Packet Loss Frequency | Badabing Tool [43] |
| Latency | Correlation between Total Runtime and Communication Time | Application Suite [30] |
| Latency | TCP/UDP/IP Transfer Delay (s, ms) | CARE [45] |
| Latency | TCP/UDP/IP Transfer Delay (s, ms) | Ping [5] |
| Latency | TCP/UDP/IP Transfer Delay (s, ms) | Send 1 byte data [20] |
| Latency | TCP/UDP/IP Transfer Delay (s, ms) | Latency Sensitive Website [5] |
| Latency | TCP/UDP/IP Transfer Delay (s, ms) | Badabing Tool [43] |
| Latency | MPI Transfer Delay (s, μs) | HPCC: b_eff [42] |
| Latency | MPI Transfer Delay (s, μs) | Intel MPI Bench [18] |
| Latency | MPI Transfer Delay (s, μs) | mpptest [8] |
| Latency | MPI Transfer Delay (s, μs) | OMB-3.1 with MPI [44] |
| Reliability | Connection Error Rate | CARE [45] |
| Reliability | Probe Loss Rate | Badabing Tool [43] |
| Data Throughput | TCP/UDP/IP Transfer bit/Byte Speed (bps, Mbps, MB/s, GB/s) | iperf [5] |
| Data Throughput | TCP/UDP/IP Transfer bit/Byte Speed (bps, Mbps, MB/s, GB/s) | Private tools TCPTest/UDPTest [43] |
| Data Throughput | TCP/UDP/IP Transfer bit/Byte Speed (bps, Mbps, MB/s, GB/s) | SPECweb 2005 [22] |
| Data Throughput | TCP/UDP/IP Transfer bit/Byte Speed (bps, Mbps, MB/s, GB/s) | Upload/Download/ Send large size data [23] |
| Data Throughput | MPI Transfer bit/Byte Speed (bps, MB/s, GB/s) | HPCC: b_eff [42] |
| Data Throughput | MPI Transfer bit/Byte Speed (bps, MB/s, GB/s) | Intel MPI Bench [18] |
| Data Throughput | MPI Transfer bit/Byte Speed (bps, MB/s, GB/s) | mpptest [8] |
| Data Throughput | MPI Transfer bit/Byte Speed (bps, MB/s, GB/s) | OMB-3.1 with MPI [44] |

*2) Computation Evaluation Metrics (cf. Table II):* Computation refers to the computing-intensive data/job processing in the Cloud. Note that, although coarse-grain Cloud-hosted applications are generally used to evaluate the overall performance of Cloud services (see *Subsection 5)*), the CPU-intensive applications have been particularly adopted for the specific Computation evaluation.

Brief descriptions of particular metrics in Table II:

- *Benchmark Efficiency vs. Instance Efficiency:* These two metrics both measure the real individual-instance Computation performance as a percentage of a baseline threshold. In *Benchmark Efficiency*, the baseline threshold is the theoretical peak of benchmark result, while it is the theoretical CPU peak in *Instance Efficiency*.
- *ECU Ratio:* This metric uses Elastic Compute Unit (ECU) instead of traditional FLOPS to measure the Computation performance. An ECU is defined as the CPU power of a 1.0-1.2 GHz 2007 Opteron or Xeon processor [42].
- *CPU Load:* This metric is usually used together with other performance evaluation metrics to judge bottleneck features. For example, low CPU load with maximum communication sessions indicate that data transfer on EC2 c1.xlarge instance is the bottleneck for a particular workload [22].

TABLE II. COMPUTATION EVALUATION METRICS

| Capacity | Metrics | Benchmark |
|---|---|---|
| Transaction Speed | Benchmark Efficiency (% Benchmark Peak) | HPL [42] |
| Transaction Speed | ECU Ratio (Gflops/ECU) | HPL [42] |
| Transaction Speed | Instance Efficiency (% CPU peak) | HPL [17] |
| Transaction Speed | Benchmark OP (FLOP) Rate (Gflops, Tflops) | DGEMM [30] |
| Transaction Speed | Benchmark OP (FLOP) Rate (Gflops, Tflops) | FFTE [30] |
| Transaction Speed | Benchmark OP (FLOP) Rate (Gflops, Tflops) | HPL [30] |
| Transaction Speed | Benchmark OP (FLOP) Rate (Gflops, Tflops) | LMbench [42] |
| Transaction Speed | Benchmark OP (FLOP) Rate (Gflops, Tflops) | NPB: EP [4] |
| Transaction Speed | Benchmark OP (FLOP) Rate (Gflops, Tflops) | Whetstone [39] |
| Latency | Benchmark Runtime (hr, min, s, ms) | Private benchmark/ application [6] |
| Latency | Benchmark Runtime (hr, min, s, ms) | Compiling Linux Kernel [46] |
| Latency | Benchmark Runtime (hr, min, s, ms) | Fibonacci [12] |
| Latency | Benchmark Runtime (hr, min, s, ms) | DGEMM [17] |
| Latency | Benchmark Runtime (hr, min, s, ms) | HPL [17] |
| Latency | Benchmark Runtime (hr, min, s, ms) | NPB [41] |
| Other | CPU Load (%) | SPECweb 2005 [22] |
| Other | Ubench CPU Score | Ubench [47] |

*3) Memory (Cache) Evaluation Metrics (cf. Table III):* Memory (Cache) is intended for fast access to temporarily saved data that can be achieved from slow-accessed hard drive storage. Since it could be hard to exactly distinguish the affect to performance brought by memory/cache, there are less evaluation practices and metrics for memory/cache than for other physical properties. However, in addition to normal capacity evaluation, there are some interesting metrics for verifying the memory hierarchies in Cloud services, as elaborated below.

TABLE III. MEMORY (CACHE) EVALUATION METRICS

| Capacity | Metrics | Benchmark |
|---|---|---|
| Transaction Speed | Random Memory Update Rate (MUP/s, GUP/s) | HPCC: RandomAccess [30] |
| Latency | Mean Hit Time (s) | Land Elevation Change App [13] |
| Latency | Memcache Get / Put / Response Time (ms) | Operate 1Byte / 1MB data [12] |
| Data Throughput | Memory bit/Byte Speed (MB/s, GB/s) | CacheBench [42] |
| Data Throughput | Memory bit/Byte Speed (MB/s, GB/s) | HPCC: PTRANS [30] |
| Data Throughput | Memory bit/Byte Speed (MB/s, GB/s) | HPCC: STREAM [42] |
| Memory Hierarchy | Intra-node Scaling | DGEMM [17] |
| Memory Hierarchy | Intra-node Scaling | HPL [17] |
| Memory Hierarchy | Sharp Performance Drop (increasing workload) | Bonnie [42] |
| Memory Hierarchy | Sharp Performance Drop (increasing workload) | CacheBench [42] |
| Other | Ubench Memory Score | Ubench [47] |

Brief descriptions of particular metrics in Table III:

- *Intra-node Scaling:* This metric is relatively complex. It is used to judge the position of cache contention by employing Scalability evaluation metrics (see *Subsection 6)*). To observe the scaling capacity of a service instance, the benchmark is executed repeatedly along with varying workload and the number of used CPU cores [17].

- *Sharp Performance Drop:* This metric is used to find cache boundaries of the memory hierarchy in a particular service instance. In detail, when repeatedly executing the benchmark along with gradually increasing workload, the major performance drop-offs can roughly indicate the memory hierarchy sizes [42].

*4) Storage Evaluation Metrics (cf. Table IV):* Storage of Cloud services is used to permanently store users' data, until the data are removed or the services are suspended intentionally. Compared to acessing Memory (Cache), accessing data permanently stored in Cloud services usually takes longer time.

TABLE IV. STORAGE EVALUATION METRICS

| Capacity | Metrics | Benchmark |
|---|---|---|
| Transaction Speed | One Byte Data Access Rate (bytes/s) | Download 1 byte data [38] |
| | Benchmark I/O Operation Speed (ops) | Bonnie/Bonnie++ [42] |
| | Blob/Table/Queue I/O Operation Speed (ops) | Operate Blob/ Table/Queue Data[5] |
| | Performance Rate between Blob & Table | Operate Blob & Table Data [20] |
| Availability | Histogram of GET Throughput (in chart) | Get data of 1Byte/100MB [9] |
| | Benchmark I/O Delay (min, s, ms) | BitTorrent [38] |
| | | Private benchmark/ application [6] |
| | | NPB: BT [4] |
| | Blob/Table/Queue I/O Operation Time (s, ms) | Operate Blob/ Table/Queue Data[5] |
| | Page Generation Time (s) | TPC-W [5] |
| Reliability | I/O Access Retried Rate | Download Data [38] |
| | | HTTP Get/Put [25] |
| Data Throughput | Benchmark I/O bit/Byte Speed (KB/s, MB/s) | Bonnie/Bonnie++ [42] |
| | | IOR in POSIX [44] |
| | | PostMark [7] |
| | | NPB: BT-IO [44] |
| | Blob I/O bit/Byte Speed (Mbps, Bytes/s, MB/s) | Operate Blob Data [38] |

Brief descriptions of particular metrics in Table IV:

- *One Byte Data Access Rate:* Although the unit here seems for Data Throughput evaluation, this metric has been particularly used for measuring Storage Transaction Speed. Contrasted with accessing large-size files, the performance of accessing very small-size data can be dominated by the transaction overheard of storage services [38].
- *Blob/Table/Queue I/O Operation metrics:* Although not all of the public Cloud providers specify the definitions, the Storage services can be categorized into three types of offers: Blob, Table and Queue [5]. In particular, the typical Blob I/O operations are *Download* and *Upload*; the typical Table I/O operations are *Get*, *Put* and *Query*; and the typical Queue I/O operations are *Insert*, *Retrieve*, and *Remove*.
- *Histogram of GET Throughput (in chart):* Unlike the other traditional metrics, this metric is represented as a chart instead of a quantitative number. In this case, the Histogram vividly illustrates the changing of GET Throughput during a particular period of time, which intuitively reflects the Availability of a Cloud service. Therefore, the Histogram chart here is also regarded as a special metric, and so do the other charts and tables in *Subsection 6)* and *7)*.

*5) Overall Performance Evaluation Metrics (cf. Table V):* In addition to the performance evaluations of specific physical properties, there are also a large number of evaluations of the overall performance of commercial Cloud services. We consider an overall performance evaluation metric as long as it was intentionally used for measuring the overall performance of Cloud services in the primary study.

Brief descriptions of particular metrics in Table V:

- *Relative Performance over a Baseline (rate):* This metric is usually used to standardize a set of performance evaluation results, which can further facilitate the comparison between those evaluation results. Note the difference between this metric and the metric *Performance Speedup over a Baseline*. The latter is a typical Scalability evaluation metric, as explained in *Subsection 6)*.
- *Sustained System Performance (SSP):* This metric uses a set of applications to give an aggregate measure of performance of a Cloud service [30]. In fact, we can find that two other metrics are involved in the calculation of this metric: the *Geometric Mean* of individual applications' *Performance per CPU Core* result is multiplied by the number of computational cores.
- *Average Weighted Response Time (AWRT):* By using the resource consumption of each request as weight, this metric gives a measure of how long on average users have to wait to accomplish their required work [33]. The resource consumption of each request is estimated by multiplying the request's execution time and the required number of Cloud service instances.

*6) Scalability Evaluation Metrics (cf. Table VI):* Scalability has been variously defined within different contexts or from different perspectives [20]. However, no matter under what definition, the evaluation of Cloud services' Scalability inevitably requires varying workload and/or Cloud resources. Since the variations are usually represented into charts and tables, we treat the corresponding charts and tables also as special metrics. In fact, unlike evaluating other performance properties, the evaluation of Scalability (and also Variability) normally implies comparison among a set of data that can be conveniently organized in charts and tables.

TABLE V. OVERALL PERFORMANCE EVALUATION METRICS

| Capacity | Metrics | Benchmark |
|---|---|---|
| Transaction Speed | Benchmark OP (FLOP) Rate (Mflops, Gflops, Mops) | HPL [4] |
| | | GASOLINE [48] |
| | | NPB [4] |
| | Benchmark Transactional Job Rate | BLAST [52] |
| | | Sysbench on MySQL [3] |
| | | TPC-W [29] |
| | | WSTest [49] |
| | Geometric Mean of Serial NPB Results (Mop/s) | NPB [44] |
| | Relative Performance over a Baseline (rate) | MODIS Processing [15] |
| | | NPB [4] |
| | Sustained System Performance (SSP) | Application Suite [30] |
| | Performance per Client | TPC-E [20] |
| | Performance per CPU Cycle (Mops/GHz) | NPB [4] |
| | Performance per CPU Core (Gflops/core) | Application Suite [30] |
| Availability | Histogram of Average Transaction Time | TPC-E [20] |
| Latency | Benchmark Delay (hr, min, s, ms) | Broadband/Epigenome/Montage [24] |
| | | CSFV [8] |
| | | FEFF84 MPI [48] |
| | | MapReduce App [47] |
| | | MCB Hadoop [50] |
| | | MG-RAST+BLAST [37] |
| | | MODIS Processing [15] |
| | | NPB-OMP/MPI [51] |
| | | WCD [23] |
| | | WSTest [49] |
| | Benchmark Transactional Job Delay (min, s) | BLAST [5] |
| | | C-Meter [16] |
| | | MODIS Processing [15] |
| | | SAGA BigJob Sys [40] |
| | | TPC-E [20] |
| | | TPC-W [53] |
| | Relative Runtime over a Baseline (rate) | Application Suite [30] |
| | | SPECjvm2008 [5] |
| | Average Weighted Response Time (AWRT) | Lublin99 [33] |
| Reliability | Error Rate of DB R/W | CARE [45] |
| Data Throughput | DB Processing Throughput (byte/sec) | CARE [45] |
| | BLAST Processing Rate (Mbp/instance/day) | MG-RAST + BLAST [37] |

Brief descriptions of particular metrics in Table VI:
- *Aggregate Performance & Performance Degradation/Slowdown over a Baseline:* These two metrics are often used to reflect the Scalability of a Cloud service (or feature) when the service (or feature) is requested with increasing workload. Therefore, the Scalability evaluation here is from the perspective of workload.
- *Performance Speedup over a Baseline:* This metric is often used to reflect the Scalability of a Cloud service (or feature) when the service (or feature) is requested for different amounts or capabilities of Cloud resources. Therefore, the Scalability evaluation here is from the perspective of Cloud resource.
- *Performance Degradation/Slowdown over a Baseline:* Interestingly, this metric can be intuitively regarded as an opposite one to the above metric *Performance Speedup over a Baseline*. However, it is more meaningful to use this metric to reflect the Scalability of a Cloud service (or feature) when the service (or feature) is requested to deal with different amount of workload. Therefore, the Scalability evaluation here is from the perspective of workload.
- *Parallelization Efficiency E(n):* Interestingly, this metric can be viewed as a "reciprocal" of the normal *Performance Speedup* metric. $T(n)$ is defined as the time taken to run a job with $n$ service instances, and then $E(n)$ can be calculated through $T(1)/T(n)/n$.

TABLE VI. SCALABILITY EVALUATION METRICS

| Sample | Metrics |
|---|---|
| [22] | Aggregate Performance |
| [13] | Performance Speedup over a Baseline |
| [20] | Performance Degradation/Slowdown over a Baseline |
| [23] | Parallelization Efficiency E(n)= T(1)/T(n)/n |
| [48] | Representation in Single Chart (Column, Line, Scatter) |
| [47] | Representation in Separate Charts |
| [42] | Representation in Table |

*7) Variability Evaluation Metrics (cf. Table VII):* In the context of Cloud service evaluation, Variability indicates the extent of fluctuation in values of an individual performance property of a commercial Cloud service. The variation of evaluation results can be caused by the performance difference of Cloud services at different time and/or different locations. Moreover, even at the same location and time, variation may still exist in a cluster of service instances. Note that, similar to the Scalability evaluation, the relevant charts and tables are also regarded as special metrics.

Brief descriptions of particular metrics in Table VII:
- *Average, Minimum, and Maximum Value together:* Although the three indicators in this metric cannot be individually used for Variability evaluation, they can still reflect the variation of a Cloud service (or feature) when placed together.
- *Coefficient of Variation (COV):* COV is defined as a ratio of the standard deviation (STD) to the mean of evaluation results. Therefore, this metric has been also directly represented as STD/Mean Rate [5].
- *Cumulative Distribution Function vs. Probability Density Function:* Both metrics distribute the probabilities of different evaluation results to reflect the variation of a Cloud service (or feature). In the existing works, the *Cumulative Distribution Function* is more popular, and often represents Scalability evaluation simultaneously through multiple distribution curves [9].

TABLE VII. VARIABILITY EVALUATION METRICS

| Sample | Metrics |
|---|---|
| [46] | Average, Minimum, and Maximum Value together |
| [6] | Coefficient of Variation (COV) (ratio) |
| [23] | Difference between Min & Max (%) |
| [20] | Standard Deviation with Average Value |
| [9] | Cumulative Distribution Function Chart |
| [43] | Probability Density Function (Frequency Function Chart) |
| [12] | Quartiles Chart with Median/Mean Value |
| [9] | Representation in Single Chart (Column, Line, Scatter/Jitter) |
| [12] | Representation in Separate Charts |
| [9] | Representation in Table |

## B. Economics Evaluation Metrics

Economics has been generally considered a driving factor in the adoption of Cloud Computing. According to the discussion about Cloud Computing from the view of Berkeley [35], the Economics aspect of a commercial Cloud service comprises two properties: Cost and Elasticity. Thus, we collected and arranged relevant metrics for these two properties respectively, as shown below.

*1) Cost Evaluation Metrics (cf. Table VIII):* Cost is an important and direct indicator to show how economical when applying Cloud Computing [35]. In theory, the Cost may cover a wide range of factors if moving computing to the Cloud. However, in the reviewed primary studies, we found that the current evaluation work mainly concentrated on the real expense of using Cloud services. By analyzing the contexts of the identified cost evaluation metrics, we have categorized them into seven metric types for easier distinction among the various metric names.

Brief descriptions of particular metrics in Table VIII:

- *Time-related & Performance-related Cost Effectiveness metrics:* Since a cost-effectiveness analysis can determine the cost per unit of outcome [14], the cost effectiveness metrics are generally expressed in a price-like manner. As the names suggest, the former type of metrics use time to measure the unit of outcome, while the latter type use performance.
- *Incremental Cost-Effectiveness Ratio metrics:* In contrast with abovementioned types, this metric type emphasizes the change, i.e., the metrics are typically expressed as a ratio of change in costs to the change in effects. Note that we kept the original names of the detailed metrics collected from the reviewed studies, although they may not be named precisely.
- *Cost Efficiency metrics:* According to the explanations in [19], we can find that the particular distinction between cost efficiency and cost effectiveness is that "efficiency is the ratio of output to input". Therefore, this type of metrics is usually expressed like reciprocals of the *Cost Effectiveness* metrics.

TABLE VIII. COST EVALUATION METRICS

| Type | Metrics | Benchmark |
|---|---|---|
| Monetary Expense | Component Resource Cost ($) | Montage/Broadband/ Epigenomics [24] |
| | | SPECweb2005 [22] |
| | Total Cost ($) | HPL [17] |
| | | Lublin99 [33] |
| | | MCB Hadoop [50] |
| | | Montage [54] |
| | | Parallel Job Exe [12] |
| | | SDSC Job Traces [33] |
| Time-related Cost Effectiveness | Cost over a Fixed Time ($/year, $/month, $/day, $/hour, $/second) | Dzero [38] |
| | | Land Elevation Change App [13] |
| | | TPC-W [29] |
| | | Montage/Broadband/ Epigenomics [24] |
| | Cost Per User per Month ($/user/month) | Cloudstone [3] |
| Performance-related Cost Effectiveness | FLOP Cost (cent/FLOP, $/GFLOP) | N/A [39] |
| | | HPL [17] |
| | Normalized Benchmark Task Cost (in ratio) | SPECjvm2008 [5] |
| | Price/Performance Ratio | NAMD [40] |
| | Transaction Cost ($/job, $/Mbp/instance, milli-cents/operation, M$/1000 transactions) | Dzero [38] |
| | | MG-RAST + BLAST [37] |
| | | Operate Table Data [5] |
| | | TPC-W [53] |
| | Throughput Cost (M$/WIPS) | TPC-W [29] |
| Incremental Cost-Effectiveness Ratio | Cost-Benefit Ratio ($/increased performance) | WSTest [49] |
| | Cost per Unit-Speedup ($/unit) | Land Elevation Change App [13] |
| Cost Efficiency | FLOP Rate Cost Wise (GFLOPS/$) | HPL [42] |
| | Transaction Cost Wise (sequences/$) | BLAST [52] |
| | Supported Users on a Fixed Budget (#/$) | Cloudstone [3] |
| | Available Resources on a Given Budget | N/A [39] |
| Bridge | EC2 CCI Equivalent Cost per Node-Hour ($/nd-hr) | Analysis and Calculation [41] |
| | In-House vs. Cloud FLOPS Equivalence Ratio | Whetstone [39] |
| Other | Cost Predictability (Variation of Cost/WIPS) | TPC-W [29] |

- *Bridge metrics:* The bridge metrics are not directly used for measuring the cost of Cloud services. As the type name suggests, they are normally used as bridges to contrast between costs of Cloud and in-house resources in an "apple-to-apple" manner. As such, we can conveniently make comparable calculations, for example, sustainable in-house resources on a fixed Cloud resource cost, or vice versa [39].

*2) Elasticity Evaluation Metrics (cf. Table IX):* Elasticity describes the capability of both adding and removing Cloud resources rapidly in a fine-grain manner. In other words, an elastic Cloud service concerns both growth and reduction of workload, and particularly emphasizes the speed of response to changed workload [11]. Although evaluating Elasticity of a Cloud service is not trivial [36], we considered a metric as an Elasticity-related metric as long as it measures the time of resource provisioning or releasing.

TABLE IX. ELASTICITY EVALUATION METRICS

| Type | Metrics | Benchmark |
| --- | --- | --- |
| Resource Acquisition Time | Provision (or Deployment) Time (s) | N/A [5] |
| | Boot Time (s) | N/A [5] |
| | Total Acquisition Time (s) | C-Meter [16] |
| Resource Release Time | Suspend Time (s) | A test program [20] |
| | Delete Time (s) | A test program [20] |
| | Total Release Time (s) | C-Meter [16] |
| Other | Cost and Time Effectiveness ($*hr/Instances(#)) | RSD algorithm [55] |

Brief descriptions of particular metrics in Table IX:
- *Resource Acquisition Time metrics:* Resource acquisition is to achieve extra Cloud resources to satisfy the workload growth. The total acquisition time can be divided into provision time and boot time [5]. The former is the latency between when a particular amount of Cloud resources is requested to when the resources are powered on. The latter is the latency after the resource provision and before ready to use.
- *Resource Release Time metrics:* Resource release is to return unnecessary Cloud resources to save expense when workload falls. If applicable, the total release time can be further divided into suspend time and delete time [20]. The former refers to the latency of stopping running the Cloud resources, while the latter measures the latency of removing the current deployment after the resources stop running.
- *Cost and Time Effectiveness:* This metric is not originally used for Elasticity evaluation [55]. However, it inspires a possible way to Elasticity measurement. In fact, Cloud elasticity is related not only to the resource scaling time but also to the resource charging basis [11]. For example, if holding the instance acquisition/release time constant, we can consider m1.small is the most elastic instance type in the standard category of Amazon EC2 service, because it charges on a 1-ECU-hour basis, while the other two charges on 4- and 8-ECU-hour bases respectively [16].

*C. Security Evaluation Metrics (cf. Table X)*

The security of commercial Cloud services has many dimensions and issues people should be concerned with [28, 35]. However, not many Security evaluations were reflected in the identified primary studies. Even in the limited studies, security evaluation was realized mainly by qualitative discussions. In fact, this finding also confirms the proposition from industry: Security is hard to quantify [58].

TABLE X. SECURITY EVALUATION METRICS

| Feature | Metrics | Sample |
| --- | --- | --- |
| Data Security | Is SSL Applicable | [22] |
| | General Discussion | [37] |
| | Communication Latency over SSL | [25] |
| Authentication | Discussion on SHA1-HMAC | [25] |
| Overall Security | Discussion using a Risk List | [38] |

Brief descriptions of particular metrics in Table X:
- *Communication Latency over SSL:* This metric is essentially not for Security evaluation. However, it can be used to reflect the influences of security settings on performance of Cloud services.
- *Discussion using a Risk List:* A more specific suggestion for Security evaluation of Cloud services is given in [38]: the security assessment can start with an evaluation of the involved risks. Therefore, this metric is to use a pre-identified risk list to discuss the security strategies supplied by Cloud services.

### III. APPLICATION OF THE METRICS CATALOGUE

As mentioned in the motivation of constructing this metrics catalogue, we can in turn use the established catalogue to facilitate the future work of evaluation of commercial Cloud services. Here we briefly introduce three application scenarios.

*A. Looking up Evaluation Metrics*

Intuitively, this catalogue can be used directly as a dictionary entry of metrics for Cloud services evaluation. Since the choice of appropriate metrics depends on the features to be evaluated [31], we can use particular Cloud service features as the retrieval key to quickly locate candidate evaluation metrics in this catalogue. Considering that the selection of metrics is essential in an evaluation [32], an available "dictionary" can clearly and significantly help identify suitable metrics within evaluation implementations. To further facilitate the metrics lookup process, we have stored all the metric data into a succinct lookup system, and deployed the system online through Google App Engine for convenience (http://cloudservicesevaluation.appspot.com/).

*B. Identifying Research Opportunities*

By observing the distribution of metrics in this catalogue, we have found several gaps in the current research into Cloud services evaluation which require more attention. For example, there is still a lack of effective metrics for evaluating elasticity of Cloud services, which supports the claim that Elasticity evaluation of a Cloud service is not trivial [36]. Meanwhile, it seems that there is no suitable metric yet to evaluate the security features of Cloud services. In fact, only four papers among the 46 reviewed primary studies mentioned Cloud services security, and the most popular evaluation approach seems only qualitative discussions around the security features. Therefore, the lack of suitable evaluation metrics could be one of the reasons why Security was not widely addressed in the selected publications. Such a finding also confirms the proposition that it is difficult to quantify security when benchmarking Cloud services [58]. Overall, these identified gaps essentially indicate research opportunities in the Cloud services evaluation domain.

*C. Inspiring Sophisticated Evaluation Metrics*

The identified metrics can be viewed as fundamentals to inspire and build relatively sophisticated metrics for Cloud services evaluation. In fact, by using relevant basic QoS metrics to monitor the requested Cloud resources, our colleagues have developed a Penalty Model to measure the imperfections in elasticity of Cloud services for a given workload in monetary units [11]. In other words, this Penalty Model works based on a set of predetermined SLA objectives and aforementioned preliminary metrics. For example, before applying the Penalty Model to an EC2 instance, the capacity of the instance should be first measured by looking at its CPU, memory, network bandwidth, etc.

Inspired by the overall performance evaluation metric *Sustained System Performance (SSP)*, particularly, we are planning to propose Boosting Metrics to accompany the benchmark suites. Like *SSP*, given different types of Cloud-based applications, Boosting Metrics are supposed to give aggregate and unified measure of performance of a Cloud service. As such, we hope the proposed Boosting Metrics can help connect the last mile of using benchmark suites.

## IV. CONCLUSIONS AND FUTURE WORK

The selection of metrics has been identified as being essential in evaluation of computer systems [32]. In fact, the metrics selection is the prerequisite of many other evaluation steps including benchmark selection [31]. In the context of Cloud Computing, however, we have not found any systematic discussion about evaluation metrics. Therefore, we proposed an investigation into the metrics suitable for Cloud services evaluation. Due to the lack of consensus of standard definition of Cloud Computing, it is difficult to point out the full scope of metrics in advance for evaluating different Cloud services. Hence, we adopted the SLR method to identify the existing studies on Cloud services evaluation and collect the de facto evaluation metrics in the Cloud Computing domain. According to the features to be evaluated, the collected metrics are related to three aspects of Cloud services, namely Performance, Economics, and Security. With respect to Performance, we have identified 9 evaluation metrics for communication, 7 for computation, 7 for memory, 11 for storage, 16 for overall performance, 7 for scalability, and 10 for variability. Under Economics, 18 and 7 evaluation metrics have been identified for cost and elasticity respectively. For Security, there are 5 evaluation metrics in total. By arranging these identified metrics following different Cloud service features, our proposed investigation essentially established a metrics catalogue for Cloud services evaluation, as shown in this paper. Moreover, the distribution of the collected evaluation metrics is particularly listed in Table XI.

TABLE XI. DISTRIBUTION OF EVALUATION METRICS

| Service Aspect | Property of the Aspect | Number of Metrics |
|---|---|---|
| Performance | Communication | 9 |
| | Computation | 7 |
| | Memory (Cache) | 7 |
| | Storage | 11 |
| | Overall Performance | 16 |
| | Scalability | 7 |
| | Variability | 10 |
| | **Total** | **67** |
| Economics | Cost | 18 |
| | Elasticity | 7 |
| | **Total** | **25** |
| Security | Data Security | 3 |
| | Authentication | 1 |
| | Overall Security | 1 |
| | **Total** | **5** |

Statistically, during this study, we found that the existing evaluation work overwhelmingly focused on the performance features of commercial Cloud services. Many other theoretical concerns about commercial Cloud Computing, Security in particular, had not been well evaluated yet in practice. In fact, the distribution of evaluation metrics shown in Table XI also reveals this phenomenon. Therefore, we roughly conclude that the Security evaluation is the relatively most difficult research topic in the Cloud services evaluation domain. Benefiting from the rich lessons people have learned from the performance evaluation of traditional computing systems, on the contrary, evaluating performance of commercial Cloud services seems not very tough. At last, although economics of adopting Cloud Computing covers a wide range of factors in theoretical discussions, the practices of cost evaluation are mainly limited to concerning the real expense of using Cloud services. Meanwhile, evaluating elasticity of Cloud services could be another hard research issue due to the lack of effective evaluation metrics.

Overall, the contribution of this metrics catalogue is multifold. Firstly, the catalogue can be used as a dictionary for conveniently looking up suitable metrics when evaluating Cloud services. We have deployed an online system to further facilitate the metrics lookup process. Secondly, research opportunities can be revealed by observing the distribution of the existing evaluation metrics in the catalogue. As mentioned previously, evaluating elasticity and security of commercial Cloud services would comprise a large amount of research opportunities, as they could be also tough research topics. Thirdly, by understanding the preliminary metrics, more sophisticated metrics can be developed for better implementing evaluation of Cloud services. In fact, we are now proposing Boosting Metrics to connect the last mile of using benchmark suites when evaluating commercial Cloud services. Accordingly, this metrics catalogue will be used to facilitate our future research in the area of Cloud services evaluation. Furthermore, new evaluation metrics will be gradually collected and/or developed to continually enrich this metrics catalogue.

ACKNOWLEDGMENT

This project is supported by the Commonwealth of Australia under the Australia-China Science and Research Fund.

NICTA is funded by the Australian Government as represented by the Department of Broadband, Communications and the Digital Economy and the Australian Research Council through the ICT Centre of Excellence program.